# Simulation of indirect $^{13}$C-$^{13}$C *J*-coupling tensors in diamond clusters hosting the NV center

A. Nizovtsev[1,2*], A. Pushkarchuk[2,3], S. Kuten[4], D. Michels[5], D. Lyakhov[5], N. Kargin[2], S. Kilin[1]

[1] Institute of Physics, Nat. Acad. Sci., 220072 Minsk, Belarus; *e-mails: apniz@dragon.bas-net.by; sergei_kilin@yahoo.com

[2] National Research Nuclear University "MEPhI", 115409, Moscow, Russia. e-mails: APNizovtsev@mephi.ru; ALPushkarchuk@mephi.ru; NIKargin@mephi.ru

[3] Institute of Physical and Organic Chemistry, Nat. Acad. Sci., 220072 Minsk, Belarus. e-mail: alexp51@bk.ru

[4] Institute for Nuclear Problems, Belarusian State University, 220006 Minsk, Belarus. e-mail: semen_kuten@list.ru

[5] Computer, Electrical and Mathematical Science and Engineering Division, 4700 King Abdullah University of Science and Technology (KAUST), 23955-6900,Thuwal, Saudi Arabia. e-mails: dominik.michels@kaust.edu.sa; dmitry.lyakhov@kaust.edu.sa

**Abstract:** Using the ORCA 5.0.2 software package we have simulated for a first time the full tensors $^n J_{KL}$ (K,L=X,Y,Z), describing n-bond *J*-coupling of nuclear spins $^{13}$C in H-terminated diamond-like clusters $C_{10}H_{16}$ (adamantane) and $C_{35}H_{36}$ as well as in the cluster $C_{33}[NV^-]H_{36}$ hosting the negatively charged NV$^-$ center. We found that, in addition to usually considered isotropic scalar $^n J$-coupling constant the anisotropic contributions to the $^n J$-coupling tensor are essential. We have also shown that the presence of the NV center affects the *J*-coupling characteristics, especially in the case of $^{13}$C-$^{13}$C pairs located near the vacancy of the NV center.

## 1. Introduction

In the past decade there was rapid progress in development of quantum magnetic sensing technologies based on nitrogen-vacancy (NV) color centers in diamond (see, e.g. [1,2] for recent reviews). Magnetometer based on single NV center can have nanometer-scale spatial resolution and exceptional sensitivity (up to ~Hz) allowing to detect target single $^{13}$C nuclear spins or coupled $^{13}$C-$^{13}$C pairs located within the diamond which can be used as long-lived quantum memory [3]. Moreover, NV-based magnetometer allows to distinguish non-equivalent nuclear spins (by their chemical shifts) of molecules located at diamond surface [4], thus enabling new exciting research area of single-spin nuclear magnetic resonance (NMR) for investigating important is-sues ranging from determination of molecular structures of inorganic/biological compounds up to medical imaging for therapeutic matters. In these respects, predicting of high-resolution NMR characteristics for studied spin systems is essential. Among them, those of indirect nuclear spin–spin coupling (the *J*-coupling), that arise due to second-order hyperfine interactions with electrons from chemical bond connecting nuclei, are important. Generally, a second-rank tensor $^n J_{KL}$ (K,L=X,Y,Z) is required to fully describe *J*-coupling between two nuclei [5]. However, up to the recent time most high-resolution NMR experiments were focused on measuring only isotropic scalar constant $^n J_{iso}=Sp\ ^n J_{KL}/3$ because the anisotropic parts of the *J*-tensor were averaged out to zero by fast molecular motion in solution-state NMR or fast magic-angle spinning (MAS) in solid-state experiments [6-8]. Meanwhile, in the case of crystalline solids, the constituent atoms are located in certain order determined by the crystal structure so that many important NMR interactions are orientation dependent and the information about anisotropic NMR interaction tensors became essential [5,8]. In particular, recently both the symmetric $^1 J_{iso}=(^1 J_{ZZ}+^1 J_{XX}+^1 J_{YY})/3$ (=70 Hz) and the asymmetric $\Delta^1 J= {}^1 J_{ZZ}-(^1 J_{XX}+^1 J_{YY})/2$ (=90 Hz) parts of the *J*-coupling tensor $^1 J_{KL}$ for the nearest-neighbor (N-N) $^{29}Si$ nuclear spins have been determined in the single-crystal silicon by measuring at four different crystallographic orientations relative to the applied magnetic field the NMR lineshapes which are sensitive to the value of $\Delta^1 J$. In diamond with $^{13}C$ nuclear spins, which is of main interest here, analogous experiment was done many years ago [10] but due to low sensitivity any *J*-coupling effects have not been studied there. Also, as far as we know, there were no theoretical works on the quantum-chemical calculation of the *J*-coupling characteristics of $^{13}C$ nuclear spins in diamond. To fill this gap, here we are presenting for a first time the results of simulation of full tensors $J_{KL}$ (K,L=X,Y,Z) describing the *J*-couplings of nuclear spins $^{13}C$ in small H-terminated diamond clusters as well as in the cluster hosting the NV- color center.

## 2. Methods and Materials

In principle, theoretical foundations of *J*-coupling are well-established [11-14] and there has been considerable progress in calculating the *J*-coupling characteristics for many simple molecules (see, e.g. [12,15-17]) including $^{13}C$-$^{13}C$ pairs [12,16-19]. However, previously most software packages were limited by calculation of the scalar *J*-coupling constants. Only recently has it become possible to calculate full *J*-coupling tensors. Here we have used for the purpose the lastest version 5.0.2 of the ORCA package. To model diamond crystal we used H-terminated carbon clusters. First, in order to test the opportunities of the package, we calculated the *J*-tensors for all possible pairs $^{13}C$-$^{13}C$ in the diamond-like adamantane molecule $C_{10}H_{16}$ (see Fig. 1a), for which the isotropic *J*-coupling constants $^1 J_{iso}$ for N-N nuclear spins $^{13}C$ was experimentally measured to be 31.4±0.5 Hz [20]. Having obtained the value ~29.9 Hz for them

(see below Fig. 2a) that is quite close to the above experimental one, we performed similar calculations for the H-terminated carbon cluster $C_{35}H_{36}$ (Fig 1b), as well as for the similar cluster $C_{33}[NV^-]H_{36}$ hosting the NV color center (Fig. 1c). It should be noted that the choice of such small clusters was due to the fact that, as is known [14-16], the calculations of the J-coupling characteristics are very computationally demanding for even modest size molecules.

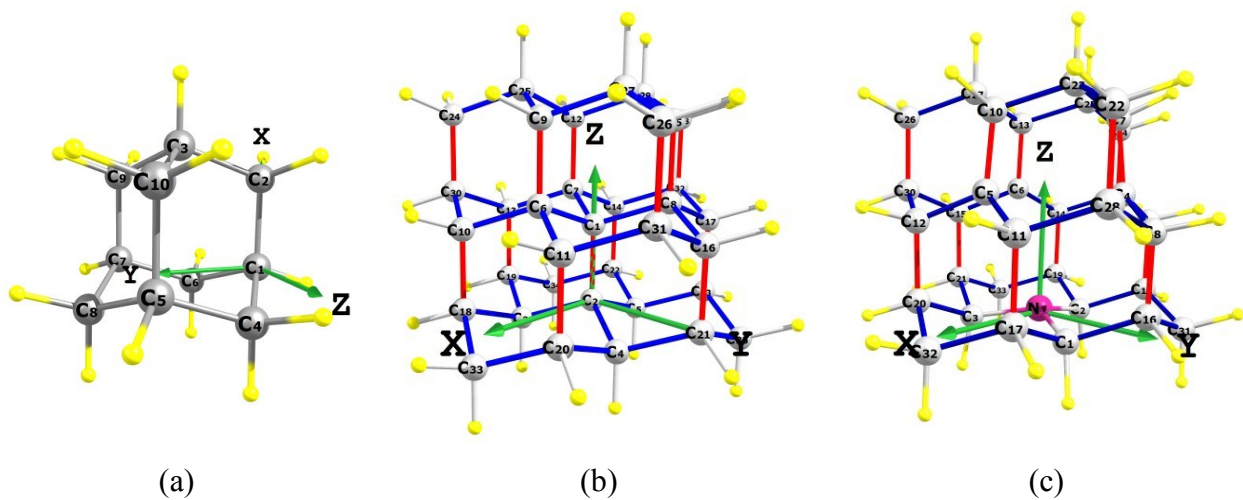

(a) (b) (c)

**Figure 1.** Simulated clusters with the carbon atoms numerated and the coordinate systems indicated. (**a**) Adamantane molecule $C_{10}H_{16}$; (**b**) Cluster $C_{35}H_{36}$; (**c**) Cluster $C_{33}[NV^-]H_{36}$. Carbon atoms Ci are shown in grey, passivating H-atoms - in yellow, nitrogen atom N in Fig. 1c – in purple.

We have optimized the cluster geometry using the ORCA 5.0.2 software package with the B3LYP/def2/J/RIJCOSX level of theory and then simulated the n-bond J-coupling tensors $^{n}J_{KL}$ for all possible $^{13}C$-$^{13}C$ pairs in the clusters using B3LYP/TZVPP/AUTOAUX/decontract level of theory. The package returns matrices describing the diamagnetic, paramagnetic, Fermi-contact, spin-dipolar and spin-dipolar/Fermi contact cross terms contributions to the total $^{n}J_{KL}$ tensor in the coordinate systems indicated in the Figure 1. Using them and taking into account the known coordinates of carbon atoms belonging to some definite $^{13}C$-$^{13}C$ pair in the cluster, one can find respective J-coupling matrices in the other coordinate system. In particular, for neighboring nuclear spins $^{13}C$, separated by a single bond in diamond (~1.54 Å), total $^{1}J_{KL}$ matrix becomes diagonal with $J_{XX} \approx J_{YY}$ in the coordinate system in which the Z-axis is directed along this bond [9]. In this case it is conventional to describe an axial J-coupling tensor in terms of two parameters: the scalar constant $^{1}J_{iso}$ and the asymmetric part $\Delta^{1}J$. Since the magnitude of the J-coupling decreases rapidly with bond order, we will mainly consider here only such N-N nuclear spins.

## 3. Results and discussion

In the case of adamantane we first calculated the isotropic J-coupling constants nJiso for all possible pairs Ci-Cj with the numbers i and j shown in the Fig. 1a. The results of calculations are shown in the Fig. 2a as a bar graph. In the molecule there are 12 pairs (C1-C2, C1-C4, C1-C6, C2-C3, C3-C4, C3-C10, C4-C5, C5-C8, C6-C7, C7-C8, C7-C9) wherein carbon atoms are nearest-neighbors. For them, the values of the $^1J_{iso}$ constants were in the range of 29.8 - 29.92 Hz, i.e. were close to the experimentally measured [20] value of 31.4 Hz. For all these pairs, the calculated total matrices $^nJ_{KL}$ were close to diagonal, since the isotropic Fermi-contact interact-

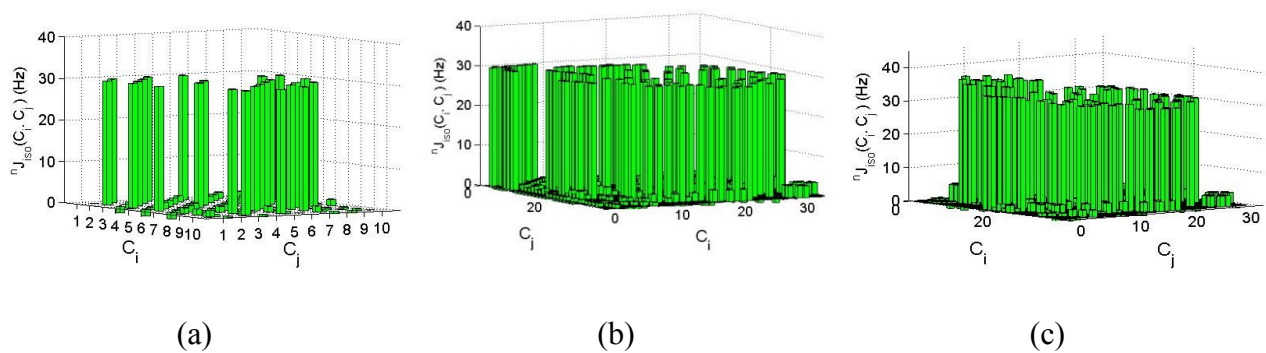

(a)            (b)            (c)

**Figure 2.** Isotropic scalar constants $^nJ_{iso}$ for all possible $^{13}C_i$-$^{13}C_j$ pairs in (**a**) adamantane molecule; (**b**) Cluster $C_{35}H_{36}$; (**c**) Cluster $C_{33}[NV^-]H_{36}$.

tion made the main contribution to them. Moreover, taking into account the symmetry of the N-N Ci-Cj pairs about their midpoint in the transformed coordinate system, in which the Z axis is directed along the Ci-Cj bond, it is possible to reduce the J-coupling matrices to their simplest diagonal form [9] with $^1J_{XX} \approx {}^1J_{YY}$. As an example, we considered the C1-C2 pair, in which both nuclear spins are located on the X axis (see Figure 1a) so that the transformation of the calculated matrices to the new coordinate system, where the Z axis is directed along the C1-C2 bond, is carried out simply by rotation counterclockwise by $90^0$ around the Y axis. For the C1-C2 pair the partial matrices in thus transformed coordinate system are presented below:

| Diamagnetic contribution (Hz): | Paramagnetic contribution (Hz): | Fermi-Contact contribution (Hz): |
|---|---|---|
| [ -0.8030    0    0.0003<br>     0    -0.8469   0.0777<br>  0.0003   -0.0576   2.5263 ]; | [ 0.2127    0    0.0002<br>    0    -0.0837   -0.0258<br>  0.0002   0.0413   -1.8121 ]; | [28.9120    0    0<br>     0    28.9120   0<br>     0      0    28.9120 ]; |

| Spin-Dipolar contribution (Hz): | SD/FC cross term contribution (Hz): | Total spin-spin $J$-coupling (Hz): |
|---|---|---|
| $\begin{bmatrix} 0.5443 & 0 & 0.0002 \\ 0 & 0.5868 & -0.0706 \\ 0.0002 & 0.0832 & 2.3375 \end{bmatrix}$ | $\begin{bmatrix} 5.0077 & 0 & -0.0015 \\ 0 & 4.9798 & -0.0605 \\ -0.0015 & -0.0605 & -9.9890 \end{bmatrix}$ | $\begin{bmatrix} 33.8736 & 0 & -0.0008 \\ 0 & 33.5480 & -0.0793 \\ -0.0008 & 0.0064 & 21.9747 \end{bmatrix}$ |

One can see from these data the relative contributions of various interactions. They also show that total matrix $^1J(C1,C2)$ is, as expected [9], near-diagonal with $^1J_{XX}(C1,C2) \approx {^1J_{YY}}(C1,C2)$ so that for this pair the asymmetric part of the $J$-coupling tensor is $\Delta^1J = {^1J_{ZZ}} - ({^1J_{XX}} + {^1J_{YY}})/2 = -11.74$ Hz. Similar data can be obtained for other pairs of N-N nuclear spins in the adamantane molecule. Figure 2a also shows that the isotropic constants $^2J_{iso}$ and 3Jiso for more distant nuclear spins are only few hertz or less.

The results of similar calculations of isotropic constants nJiso, performed for all possible pairs $^{13}C$-$^{13}C$ in the clusters $C_{35}H_{36}$ and $C_{33}[NV^-]H_{36}$ are shown in Figures 2b and 2c, respectively. As one can see from Figure 1b, for the cluster $C_{35}H_{36}$ we choose the coordinate systems in which the origin was taken at the C2 carbon atom and Z axis was directed from C2 to C1 atom. In this cluster there are 595 different $^{13}C$-$^{13}C$ pairs with 52 of them being N-N carbons. Among these N-N pairs 13 ones have their bonds being near-parallel to the chosen Z axis. These bonds are shown in red in Figure 1b. For the remaining 39 N-N pairs, shown in blue in Figure 1b, the angles between their bonds and the Z axis were approximately equal to the tetrahedral angle $109.47^0$ (or $180^0 - 109.47^0$). Respectively, for the cluster $C_{33}[NV^-]H_{36}$ the origin of the coordinate system was taken on the N atom of the NV center and Z axis was coinciding with the NV center axis. In this cluster there are 45 N-N $^{13}C$-$^{13}C$ pairs with 12 of them having bonds directed near-parallel to the Z axis. Again, these 12 pairs are shown in red in Figure 1c and the other ones are shown in blue. For the cluster $C_{35}H_{36}$ simulated one-bond constants $^1J_{iso}$ were in the range 29.8-30.0 Hz (see also Table 1 below), i.e. very close to those obtained for the admantane molecule. Contrary, in the case of the cluster $C_{33}[NV^-]H_{36}$, containing the NV center, there are several pairs of N-N $^{13}C$ atoms, located near the vacancy of the NV center, for which the values of the 1Jiso constants are slightly higher (~37.1 Hz) than for the other pairs (~ 31.5-31.8 Hz, see, in particular, Table 1).

The above data on the isotropic constants $^1J_{iso}$ for the clusters $C_{35}H_{36}$ and $C_{33}[NV^-]H_{36}$ have been obtained from total $J$-coupling matrices $^1J_{KL}$ calculated for these clusters. Generally, as in the case of adamantane, the matrices have diagonal elements being much larger than non-diagonal ones. It is these diagonal elements that are shown in Figures 3a-c and 3d-f for the clusters $C_{35}H_{36}$ and $C_{33}[NV^-]H_{36}$, respectively. In these figures, the red bars give the values of the corresponding diagonal elements for those adjacent carbon pairs for which the C-C bond is directed almost parallel to the Z axis of the coordinate system used, while the blue bars are for pairs for which the C-C bond makes a tetrahedral angle with the Z axis. More specifically, the values of the diagonal elements $^1J_{KK}$ of the $J$-coupling matrices of N-N $^{13}C$-$^{13}C$ pairs shown in red in Figures 3 are given below in Table 1 along with the corresponding values of the isotropic constants shown in Figures 2b and 2c.

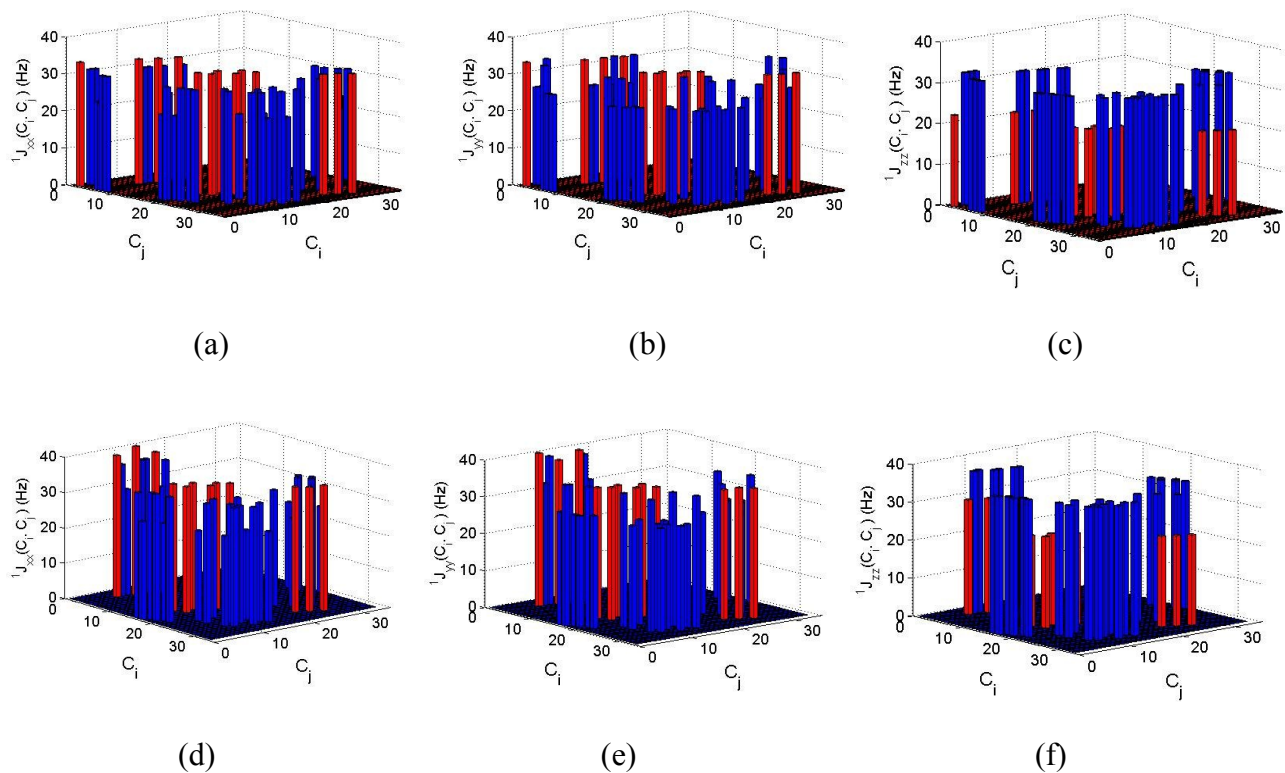

**Figure 3.** Diagonal elements $^1J_{XX}(C_i,C_j)$ **(a,d)**; $^1J_{YY}(C_i,C_j)$ **(b,e)**; and $^1J_{ZZ}(C_i,C_j)$ **(c,f)** of the $^1J$-coupling matrices, calculated for nearest-neighbor $^{13}C$-$^{13}C$ pairs in the $C_{35}H_{36}$ cluster (Figures 3a-3c) and in the $C_{33}[NV^-]H_{36}$ cluster (Figures 3d-3f).

**Table 1.** Diagonal elements $^1J_{KK}$ of total $J$-coupling tensors and corresponding values of the isotropic constants $^1J_{iso}$ calculated for N-N $^{13}C$-$^{13}C$ pairs having their bonds being near-parallel to the Z-axis of the coordinate systems shown in Figures 1b,c. Left panel shows the data for the cluster $C_{35}H_{36}$, right one – for the cluster $C_{33}[NV^-]H_{36}$.

Cluster $C_{35}H_{36}$

| Ci-Cj | $^1J_{xx}$ | $^1J_{yy}$ | $^1J_{zz}$ | $^1J_{iso}$ |
|---|---|---|---|---|
| C2-C1 | 33.45 | 33.45 | 22.41 | 29.77 |
| C6-C9 | 33.66 | 33.44 | 22.52 | 29.87 |
| C7-C12 | 33.49 | 33.60 | 22.51 | 29.87 |
| C8-C15 | 33.49 | 33.60 | 22.51 | 29.87 |

Cluster $C_{33}[NV^-]H_{36}$

| Ci-Cj | $^1J_{xx}$ | $^1J_{yy}$ | $^1J_{zz}$ | $^1J_{iso}$ |
|---|---|---|---|---|
| **C4-C7** | **39.93** | **41.41** | **30.08** | **37.14** |
| **C5-C10** | **42.06** | **39.12** | **30.01** | **37.06** |
| **C6-C13** | **39.93** | **41.41** | **30.08** | **37.14** |
| C8-C16 | 35.88 | 35.50 | 23.99 | 31.79 |

| | | | | | | | | | |
|---|---|---|---|---|---|---|---|---|---|
| C18- C10 | 32.71 | 32.79 | 21.56 | 29.02 | C11-C17 | 35.68 | 35.71 | 24.02 | 31.80 |
| C20- C11 | 32.71 | 32.79 | 21.57 | 29.02 | C9-C18 | 35.53 | 35.89 | 24.00 | 31.81 |
| C19- C13 | 32.86 | 32.67 | 21.58 | 29.04 | C14-C19 | 35.53 | 35.89 | 24.00 | 31.81 |
| C22- C14 | 32.71 | 32.82 | 21.58 | 29.04 | C12-C20 | 35.68 | 35.71 | 24.02 | 31.80 |
| C21- C16 | 32.85 | 32.66 | 21.57 | 29.03 | C15-C21 | 35.88 | 35.50 | 23.99 | 31.79 |
| C23- C17 | 32.71 | 32.82 | 21.58 | 29.03 | C22-C28 | 35.41 | 35.18 | 23.79 | 31.46 |
| C30- C24 | 32.44 | 32.32 | 20.90 | 28.55 | C24-C29 | 35.07 | 35.53 | 23.80 | 31.47 |
| C31- C26 | 32.44 | 32.31 | 20.90 | 28.55 | C26-C30 | 35.40 | 35.18 | 23.79 | 31.46 |
| C32- C28 | 32.254 | 32.513 | 20.91 | 28.56 | | | | | |

One can see from Figures 3 and, especially, from Table 1 that for the Ci-Cj pairs being near-parallel to the Z axis the values $^1J_{XX}(C_i,C_j) \approx ^1J_{YY}(C_i,C_j)$ are about one and a half times more larger then $^1J_{ZZ}(C_i,C_j)$. Moreover, the presence of the negatively charged NV- center in the cluster $C_{33}[NV^-]H_{36}$, which introduces additional electron density, leads to some increase in the diagonal elements $^1J_{KK}$ of the J-coupling matrices for all Ci-Cj pairs compared to the cluster $C_{35}H_{36}$. As follows from Table 1, such an increase in the $^1J_{KK}$ values is especially pronounced (~9%) for the C4-C7, C5-C10 and C6-C13 pairs, in which the atoms C4,C5 and C6 are the nearest-neighbors of the vacancy of the NV center and on which the electron density of the center is mainly localized [21]. A similar increase in J-coupling takes place for other pairs C4/5/6-Cj, for which the corresponding bonds make an angle of ~109.47$^0$ with the axis Z of the selected coordinate system.

**Conclusions**

For the first time, the total tensors describing the indirect interaction of $^{13}C$ nuclear spins in adamantane molecules, H-terminated diamond cluster $C_{35}H_{36}$, and in the cluster $C_{33}[NV^-]H_{36}$ hosting NV centers, have been calculated by quantum chemistry methods. It is shown that the presence of the NV center leads to a change in characteristics of the indirect interaction of $^{13}C$ nuclear spins. The results obtained are important for quantum information and sensor applications, in particular, for creation of long-lived quantum memory based on singlet-state $^{13}C$-$^{13}C$ dimers in diamond [3], creation of nanoscale NV-based quantum sensors that ensure the detection of adsorbed molecules/radicals on the surface of nanostructured diamond [4] and the determination of their chemical structure. Such sensors can be used to study biological processes at the level of individual cells, membranes, nerve fibers, targeted drug delivery and control of such delivery. The data obtained can also be useful for studies of NMR in the Zero- to Ultralow-field regime [22-25] where precisely the internal spin interactions in their natural environment are dominated.

An analysis of the dynamics of multispin systems $^{14}$NV-$^{13}$C-$^{13}$C with the account of both direct dipole-dipole interactions of $^{13}$C nuclear spins and of their J-coupling, as well as modeling of the NMR spectra of such objects, will be presented elsewhere.

**Author Contributions:** Conceptualization, A.N. and A.P.; methodology, A.N., A.P. and S.Ku.; software, A.P., D.M. and D.L.; validation, N.K. and S.Ki.; formal analysis, A.N.; investigation, A.N., A.P. and S.Ku.; resources, D.M. and D.L.; data curation, A.P. and S.Ku.; writing—original draft preparation, A.N.; writing—review and editing, A.N.; visualization, A.N. and N.K.; supervision, N.K. and S.Ki.; project administration, A.N. and N.K.; funding acquisition, A.N. and N.K. All authors have read and agreed to the published version of the manuscript.

**Funding:** This research was funded by RSF-DFG, project № 21-42-04416, and, in the part of calculations for adamantine, by the Belarus State Scientific Program Convergence-2025.
**Institutional Review Board Statement:** This study did not require ethical approval.
**Data Availability Statement:** Not applicable.
**Acknowledgments:** All Orca 5.0.2 package computations were performed on KAUST's Ibex HPC. The authors thank the KAUST Supercomputing Core Lab team for assistance with execution tasks on Skylake nodes. We are also grateful to F. Jelezko for very useful cooperation.
**Conflicts of Interest:** The authors declare no conflict of interest.

**References**
1. Schwartz, I.; Rosskopf, J.; Schmitt, S.; Tratzmiller, B.; Chen, Q.; McGuinness, L.P.; Jelezko, F.; Plenio, M.B. Blueprint for nanoscale NMR. *Scientific Reports*. **2019**, *9*, 6938.
2. Barry, J.F.; Schloss, J.M.; Bauch, E.; Turner, M.J.; Hart, C.A.; Pham, L.M.; Walsworth, R.L. Sensitivity optimization for NV-diamond magnetometry. *Rev. Mod. Phys.* **2020**, *92*, 015004.
3. Chen, Q.; Schwarz, I.; Plenio, M.B. Steady-state preparation of long-lived nuclear spin singlet pairs at room temperature. *Phys. Rev. B*. **2017,** *95*, 224105.
4. Glenn, D.R. ; Bucher, D.B.; Lee, J.; Lukin, M.D.; Park, H.; Walsworth, R.L. High-resolution magnetic resonance spectroscopy using a solid-state spin sensor. *Nature*. **2018,** *95,* 55535.
5. Harris, K.J.; Bryce, D.L.; Wasylishen, R.E. NMR line shapes from AB spin systems in solids – The role of antisymmetric spin-spin coupling. *Can. J. Chem.* **2009**, *87,* 1338-1351.
6. Frydman, L. Spin-1/2 and beyond: A perspective in solid state NMR spectroscopy. *Annu. Rev. Phys. Chem.* **2001,** *52,* 463–98.

7. Reif, B.; Ashbrook, Sh.E.; Emsley, L.; Hong, M. Solid- state NMR spectroscopy. *Nature Reviews. Methods Primers*. 2021. Article citation ID: (2021) 1:2.
8. Vaara, J.; Jokisaari, J.; Wasylishen, R.E.; Bryce, D.L. Spin–spin coupling tensors as determined by experiment and computational chemistry. *Progress in Nuclear Magnetic Resonance Spectroscopy*. **2002,** *41*, 233–304.
9. Christensen, B.; Price, J.C. NMR lineshape of $^{29}$Si in single-crystal silicon. *Phys. Rev. B*. **2017,** *95*, 134417.
10. Lefmann, K.; Buras, B.; Pedersen, E.J.; Shabanova, E.S.; Thorsen, P.A.; Rasmussen, F.B.; Sellschop, J.P.F. NMR spectra of pure $^{13}$C diamond. *Phys. Rev B*, **1994**. *50*, 15623-15627.
11. Ramsey, F. Electron Coupled Interactions between Nuclear Spins in Molecules. *Phys. Rev*. **1953**, *91*, 303-307.
12. Wray, V. Carbon-carbon coupling constants: A compilation of data and a practical quide. *Progress in NMR Spectroscopy*. **1979**, *13*, 177-256.
13. Helgaker, T.; Jaszunski, M.; Ruud, K. Ab Initio Methods for the Calculation of NMR Shielding and Indirect Spin−Spin Coupling Constants. *Chem. Rev*. **1999**, *99*, 293–352.
14. Helgaker, T.; Jaszuński, M.; Pecul, M. The quantum-chemical calculation of NMR indirect spin–spin coupling constants. *Progress in Nuclear Magnetic Resonance Spectroscopy*. **2008,** *53*, 249–268.
15. Antušek, A.; Kędziera, D.; Jackowski, K.; Jaszuński, M.; Makulski, W. Indirect spin–spin coupling constants in $CH_4$, $SiH_4$ and $GeH_4$ – Gas-phase NMR experiment and ab initio calculations. *Chemical Physics*. **2008**, *352*, 320–326.
16. Krivdin, L.B.; Contreras, R.H. Recent Advances in Theoretical Calculations of Indirect Spin–Spin Coupling Constants. *Annu. Rep. Nucl. Magn. Reson.Spectrosc*. **2007**, *61*, 133–245.
17. Kamienska-Trela, K. One-bond $^{13}$C-$^{13}$C Spin-Spin Coupling Constants. *Annual reports on NMR spectroscopy*. **1995**, *30*, 131-222.
18. Jaszunsi, M.; Ruud, R.; Helgaker, T. Density-functional theory calculation of the nuclear magnetic resonance indirect nuclear spin—spin coupling constants in $C_{60}$. *Mol. Phys*. **2003**, *101*, 1997-2002.
19. Peralta, J.E.; Barone, V.; Scuseria, G.E; Contrera, R.H. Density Functional Theory Calculation of Indirect Nuclear Magnetic Resonance Spin-Spin Coupling Constants in $C_{70}$. *J. Am. Chem. Soc*. **2004**, *126*, 7428-7429.
20. Gay, I.D.; Jones, C.H.W.; Sharma R. D. INADEQUATE in the Solid State. Homonuclear Couplings in [ ( $CH_3$)2SnE]$_3$. *J. of Magnetic Resonance*. **1991**, *91*, 186-189.

21. Nizovtsev A.P.; Kilin S.Ya.; Pushkarchuk, A.L.; Pushkarchuk,V.A.; Kuten, S.A.; Zhikol, O.A.; Schmitt, S.; Unden, T.; Jelezko, F. Non-flipping $^{13}$C spins near an NV center in diamond: hyperfine and spatial characteristics by density functional theory simulation of the $C_{510}[NV]H_{252}$ cluster. New J. Phys. **2018**, *20*, 023022.
22. Theis, T.; Blanchard, J.W.; Butler, M.C.; Ledbetter, M.P.; Budker, D.; Pines, A. Chemical analysis using J-coupling multiplets in zero-field NMR. *Chemical Physics Letters.* **2013**, *580*, 160–165.
23. Blanchard, J.W.; Budker, D. Zero- to Ultralow-field NMR. *eMagRes.* **2016,** *5*, 1395–1410.
24. Jiang, M.; Wu, T.; Blanchard, J.W.; Feng, G.; Peng, X.; Budker, D. Experimental benchmarking of quantum control in zero-field nuclear magnetic resonance. *Sci. Adv.* **2018**, *4*, eaar6327, 1-7.
25. DeVience, S.J.; Greer, M.; Mandal, S.; Rosen, M.S. Homonuclear *J*-Coupling Spectroscopy at Low Magnetic Fields using Spin-Lock Induced Crossing. *Chem Phys Chem*. **2021**, *22*, 2128-2137.